\documentclass[10pt,aps,prl,twocolumn,showpacs,preprintnumbers,amsmath,amssymb,unsortedaddress,floatfix]{revtex4-1}
\usepackage{graphicx}  %  Include figure files
\usepackage{dcolumn}   % Align table columns on decimal point
\usepackage{bm}   
\usepackage{natbib}

\newcommand\Eqn[1]     {Eq.\,(\ref{#1})}

\newcommand\nn         {\nonumber}

\def\mnras{{Mon.~ Not.~ R.~ Astron.~ Soc.~}}

\def\prd{{Phys.~ Rev.~ D.~}}

\def\apj{{Astrophys.~ J.~}}

\def\nat{{Nature (London)~}}

\def\mnras{{MNRAS}}

\def\prd{{PRD}}

\def\apj{{ApJ}}

\def\aap{{A\&A}}
\def\nat{{Nature}}

\def\physrep{{Phys.~ Rep.~}}

\newcommand{\be}{\begin{equation}}
\newcommand{\ee}{\end{equation}}
\newcommand{\ba}{\begin{eqnarray}}
\newcommand{\ea}{\end{eqnarray}}

\def\pp1{{\prime}}
\def\pp2{{\prime\prime}}

\def\2D{{\rm 2D}}

\def\SN{{\mathcal S}/{\mathcal N}}

\def\g{{\rm g}}

\def\s{{\rm s}}

\def\bx{{\bf x}}
\def\br{{\bf r}}

\def\bk{{\bf k}}
\def\bq{{\bf q}}

\def\1Loop{{\rm 1Loop}}

\def\Msol{h^{-1}M_{\odot}}

\def\Mpc{\, h^{-1}{\rm Mpc}}

\def\kMpc{\, h \, {\rm Mpc}^{-1}}

\def\dr{d^3r}
\def\dx{d^3x}
\def\dk{d^3k}
\def\dq{d^3q}

\def\nbar{\bar{n}}
\def\nbarg{\bar{n}_{\rm g}}

\def\dirac{\delta^{\rm D}}

\def\fun#1#2{\lower3.6pt\vbox{\baselineskip0pt\lineskip.9pt
        \ialign{$\mathsurround=0pt#1\hfill##\hfil$\crcr#2\crcr\sim\crcr}}}

\def\Nh{{N_{\rm h}}}
\def\Ng{{N_{\rm g}}}

\def\Fg{{\mathcal F}_{\rm g}}
\def\Fgt{\tilde{\mathcal F}_{\rm g}}
\def\Gb{\overline{\mathcal G}}
\def\G{\mathcal {G}}
\def\Gt{\tilde{\mathcal {G}}}

\def\Lmin{L_{\rm min}}

\def\pvp{\mbox {\rm\tiny PVP}}
\def\fkp{\mbox {\rm\tiny FKP}}
\def\opt{\mbox {\rm\tiny OPT}}
\def\SN{{\mathcal S}/{\mathcal N}}

\def\nbarg{\overline{n}_{\g}}

\def\bLSqbar{\overline{b_{L}^2}}
\begin{document}
%%%%%%%%%%%%%%%%%%%%%%%%%%%%%%%%%%%%%%%%%%%%%%%%%%%%%%%
\title{What is the optimal way to measure the galaxy power spectrum?}
%%%%%%%%%%%%%%%%%%%%%%%%%%%%%%%%%%%%%%%%%%%%%%%%%%%%%%%
\author{Robert E.~Smith$^{1}$} 
\email[E-mail me at: ]{r.e.smith@sussex.ac.uk}
\author{Laura Marian$^{1}$}
\affiliation{(1) Department of Physics and Astronomy, 
University of Sussex, Brighton BN1 9QH, UK}

%%%%%%%%%%%%%%%%%%%%%%%%%%%%%%%%%%%%%%%%%%%%%%%%%%%%%%%
\begin{abstract}
Measurements of the galaxy power spectrum contain a wealth of
information about the Universe.  Its optimal extraction is vital if we
are to truly understand the micro-physical nature of dark matter and
dark energy.  In Smith \& Marian (2015) we generalized the power
spectrum methodology of Feldman et al. (1994) to take into account the
key tenets of galaxy formation: galaxies form and reside exclusively
in dark matter haloes; a given dark matter halo may host galaxies of
various luminosities; galaxies inherit the large-scale bias associated
with their host halo.  In this paradigm we derived the optimal
weighting and reconstruction scheme for maximizing the signal-to-noise
on a given band power estimate. For a future all-sky flux-limited
galaxy redshift survey of depth $b_{\rm J} >22$, we now demonstrate
that the optimal weighting scheme does indeed provide improved $\SN$
at the level of $\sim20\%$ when compared to Feldman et al. (1994) and
$\sim60\%$ relative to Percival et al. (2003), for scales of order
$k\sim0.5\kMpc$. Using a Fisher matrix approach, we show that the
cosmological information yield is also increased relative to these
alternate methods -- especially the primordial power spectrum
amplitude and dark energy equation of state. 
\end{abstract}

%%%%%%%%%%%%%%%%%%%%%%%%%%%%%%%%%%%%%%%%%%%%%%%%%%%%%%%
\maketitle
%%%%%%%%%%%%%%%%%%%%%%%%%%%%%%%%%%%%%%%%%%%%%%%%%%%%%%%
\noindent
{\bf{\em Introduction}} --- The matter power spectrum is a fundamental tool
for constraining the cosmological parameters. It contains detailed
information about the large-scale geometrical structure of space-time,
as well as the phenomenological properties of dark energy and dark
matter.  Given a galaxy redshift survey two things are crucial: how to
obtain an {\it unbiased} and {\it optimal} estimate of the information
in the matter fluctuations.

State-of-the-art galaxy redshift surveys, such as the Baryon
Oscillation Spectroscopic Survey \citep[][hereafter
  BOSS]{Andersonetal2012,Andersonetal2014a,Andersonetal2014b}, Galaxy
And Mass Assembley \citep[][hereafter GAMA]{Blakeetal2013}, and
WiggleZ \citep{Blakeetal2011}, have all used the approach of
\citet[][hereafter FKP]{Feldmanetal1994} to estimate the power
spectrum. This assumes that galaxies are a Poisson sampling of the
underlying density field. Hence, provided one subtracts an appropriate
shot-noise term, and deconvolves for the survey window function, one
should obtain an unbiased estimate of the matter power spectrum.

In the last two decades, our understanding of galaxy formation has
made rapid progress since the work of FKP and our current best models
strongly suggest that galaxies are not related to matter in the way
they envisioned
\citep{WhiteRees1978,WhiteFrenk1991,Kauffmannetal1999,Bensonetal2000,Springeletal2005}.
Furthermore, observational studies have discovered that galaxy
clustering depends on various physical properties: e.g. luminosity
\citep{Parketal1994,Norbergetal2001,Norbergetal2002a,Zehavietal2002,Zehavietal2005,Swansonetal2008,Zehavietal2011},
colour
\citep{Brownetal2000,Zehavietal2002short,Zehavietal2005short,Swansonetal2008,Zehavietal2011short}
morphology \citep{DavisGeller1976,Guzzoetal1997,Norbergetal2002a}, and
stellar mass \citep{Lietal2006} etc.

\citet[][hereafter PVP]{Percivaletal2004a} attempted to correct the
FKP framework to take into account the effects of luminosity-dependent
clustering. In a recent paper \citep[][hereafter
  SM15]{SmithMarian2015}, we argued that the approach of PVP, whilst
appearing qualitatively reasonable, is in fact at odds with our
current understanding of galaxy formation, and so non-optimal. More
recent studies by \citet{Seljaketal2009}, \citet{Hamausetal2010} and
\citet{Caietal2010} suggested that weighting the galaxy density field
by a linear function of halo mass would reduce stochasticity.

In SM15 we developed a new scheme incorporating a number of the key
ideas from galaxy formation: galaxies only form in dark matter haloes
\citep{WhiteRees1978}; haloes can host galaxies of various
luminosities; the large-scale bias associated with a given galaxy is
largely inherited from the bias of the host dark matter halo.
 
In this work we demonstrate that our new optimal estimator indeed
provides both improved signal-to-noise (hereafter $\SN$) estimates of
the galaxy power spectrum and boosted cosmological information
content, when compared with the FKP and PVP approaches.

This letter is broken down as follows: First, we provide a brief
overview of the results from SM15. Next, we evaluate the $\SN$
expressions for the various weighting schemes, followed by the
cosmological information from a putative all-sky galaxy redshift
survey. Finally, we conclude.
 
%%%%%%%%%%%%%%%%%%%%%%%%%%%%%%%%%%%%%%%%%%%%%%%%%%%%%%%
\vspace{0.2cm}
\noindent
{\bf{\em Optimal power spectrum estimation}} --- In the original work of
FKP, the starting concept is that galaxies are simply an independent
point sampling of the underlying galaxy field. Hence,
\vspace{-0.4cm}
\ba 
n_{\g}(\br) & = & \sum_{i=1}^{\Ng}\dirac(\br-\br_i)\label{eq:ng_FKP} \ ,
\ea
where $N_g$ is the number of galaxies, and $\br_i$ is the position of
the $i$th galaxy in the survey. From this field one then may construct
an effective galaxy over-density field:
\ba 
\Fg^{\fkp}(\br) =\Theta(\br) w(\br)\left[\frac{}{}n_{\g}(\br)-\alpha n_{\s}(\br)\right],
\label{eq:gal_over_FKP}
\ea
where $\Theta(\br)$ is a survey mask function, which is 1 if the
galaxy lies inside the survey volume and 0 otherwise, $\alpha$ is a
scaling factor for the spatially random galaxy field $n_{\s}(\br)$ and
$w(\br)$ is an optimal weight function that depends on $\br$.  If we
now follow the FKP logic and compute the power spectrum of the
$\Fg^{\fkp}$ field, one finds that it is related to the galaxy power
spectrum $P_{\rm g}(k)$ through the relation:
\be \left<|\Fg^{\fkp}(\bk)|^2\right> =
\int \frac{\dk'}{(2\pi)^3}P_{\rm g}(\bk')\left|\Gt^{\fkp}(\bk-\bk')\right|^2
+P^{\fkp}_{\rm shot}\label{eq:Pf} \ ,\ee
where $\Gt^{\fkp}(\bk)$ is the weighted version of the Fourier
transform of the survey mask function $\Theta(\br)$, and
$P^{\fkp}_{\rm shot}$ is an effective shot noise correction. If one
subtracts the shot noise and deconvolves for the survey window
function, then one may obtain an estimate of $P_{\rm g}(k)$. It is
important to realise that the above procedure is only correct under
the assumption that the galaxy power spectrum does not depend on any
observable, e.g. galaxy luminosity, colour, spectral type, host halo
mass, etc. and that shot noise is as was given by FKP. If these
assumptions are wrong, then the functions $P_{\rm g}(k)$,
$\Gt^{\fkp}(\bk)$, and $P^{\fkp}_{\rm shot}$ will all pick up these
dependencies, resulting in a biased and sub-optimal reconstruction of
the `true' power spectrum \footnote{This bias is lessened if the
  survey is volume limited, but the bias persists in the form of a
  modification to the shot noise.}.  As noted earlier, current
observational evidence indicates that clustering strength does depend
on the sample selection. Hence, FKP must be biased and sub-optimal
(PVP also came to a similar conclusions).

We now summarise the SM15 formalism, designed to account for a number
of these effects. Consider a large survey volume containing $N_g$
galaxies that are constrained to be distributed inside $N_h$ dark
matter haloes. Thus the $i$th dark matter halo of mass $M_i$ and
position of the centre of mass $\bx_i$, will have $N_\g(M_i)$
galaxies. The $j$th galaxy will have a position vector $\br_j$
relative to the centre of the halo and a luminosity $L_j$.  For this
more complicated distribution, \Eqn{eq:ng_FKP} can be generalized to:
\ba 
n_{\g}(\br,L,\bx,M) & = & \sum_{i=1}^{\Nh}\dirac(\bx-\bx_i)\dirac(M-M_i) \nn \\
& \times & \sum_{j=1}^{\Ng(M_i)}\dirac(\br-\br_j-\bx_i)\dirac(L-L_j),
\label{eq:ng_def}
\ea
where the four Dirac delta functions, going from right to left, are
sampling: the luminosity of each galaxy in a given halo; the spatial
location of a given galaxy relative to the halo centre; the halo mass
from the distribution of masses; and the halo centre in the survey
volume. 

In direct analogy with FKP's \Eqn{eq:gal_over_FKP}, we define an
effective galaxy over-density field:
\ba 
\Fg(\br) &=& \int dL \int \dx
\int dM \Theta(\br|L)\frac{w(\br,L,\bx,M)}{\sqrt{A}} \nn \\
& & \hspace{-0.2cm} \times \left[\frac{}{}n_{\g}(\br,L,\bx,M)-\alpha n_{\s}(\br,L,\bx,M)\right],
\label{eq:gal_over} 
\ea
where $\Theta(\br|L)$ is the luminosity-dependent survey geometry
function; $A$ is a normalisation constant; $\alpha$ is a scaling
factor for the random halo catalogue; $w$ is a general weight
function.  The function $n_{\s}$ is the same as $n_{\g}$, except the
spatial locations of the halo centres have been randomized and the
number density has been scaled up by a factor of $1/\alpha$.  As shown
by SM15, in the large-scale limit, when the distribution of galaxies
in haloes adopts a Dirac-delta-function-like behaviour, the $\Fg$
power spectrum can be written:
\be
\left<|\Fgt(\bk)|^2\right>  \approx  \int \frac{\dq}{(2\pi)^3} P(\bq) \left|
\Gt^{(1)}_{(1,1)}(\bk-\bq)\right|^2 + P_{\rm shot}  \ ,
\label{eq:PhmLS}
\ee
where $P(\bq)$ is the true {\em matter} power spectrum, which is
convolved with the effective survey window function
$\Gt^{(1)}_{(1,1)}(\bk)$, and $P_{\rm shot}$ is a new effective shot
noise. The set of survey window functions that are required to
evaluate these expressions can be written in general:
\ba \G^{(n)}_{(l,m)}(\br) & \equiv & A^{-n l/2}\hspace{-0.2cm} \int dM \nbar(M,\chi)
b^{m}(M,\chi) N^{(n)}_{\g}(M) \nn \\
& & \hspace{-1cm}\times\left[\int dL \Theta(\br|L)\Phi(L|M)w^l(\br,L,\bx,M)\right]^n \ ,
\label{eq:G} \ea
where in the above $\nbar(M,\chi)$ and $b(M,\chi)$ are the mass
function and large-scale bias of haloes of mass $M$ at radial position
$\chi$ from the observer ($\chi$ here is also acting as coordinate
time); $N^{(n)}_{\g}(M)$ gives the $n$th factorial moment of the halo
occupation distribution (hereafter HOD); $\Phi(L|M)$ gives the
conditional probability density that a galaxy hosted in a halo of mass
$M$ has a luminosity $L$.  Using these functions the effective shot
noise term can be written:
\be
P_{\rm shot} \equiv
(1+\alpha)\left[\int\frac{\dq}{(2\pi)^3} \Gt^{(2)}_{(1,0)}({\bq})
+ \Gt^{(1)}_{(2,0)}({\bf 0})\right] \ ,
\label{eq:Pshot} \ee
We also introduce the normalisation-free window functions
$\Gb^{(n)}_{(l,m)}=A^{n l/2} \G^{(n)}_{(l,m)}$, which enables us to
write:
$ A \equiv  \int \dr \left|\Gb^{(1)}_{(1,1)}(\br) \right|^2 .
%\label{eq:norm_def} \ .
$
We thus see that, similar to the FKP approach, in order to recover the
matter power spectrum, one must subtract the effective shot-noise term
and deconvolve for the square of the effective survey window function
$\Gt^{(1)}_{(1,1)}(\bk)$.

In the large-scale limit and under the assumption that the matter
density field is Gaussianly distributed, SM15 also showed that the
$\SN$ can, for an arbitrary weight $w$, be written in general as:
\begin{widetext}
\ba
\left(\frac{\mathcal S}{\mathcal N}\right)^2 = 
\frac{V(k)}{2 (2\pi)^3}\;\left[\int \dr \left|\Gb^{(1)}_{(1,1)}(\br)\right|^2\right]^2 
\left\{\int \dr \left(\left[\Gb^{(1)}_{(1,1)}(\br)\right]^2 + 
\frac{(1+\alpha)}{\overline{P}(k)} \left[ \Gb^{(2)}_{(1,0)}(\br) + \Gb^{(1)}_{(2,0)}(\br)\right]
\right)^2\right\}^{-1} \ ,
\label{eq:SN_gen}
\ea
\end{widetext}
where $V(k)=4\pi k_i^2 \Delta k \left[1+\left({\Delta
    k}/{k_i}\right)^2/12\right]$ is the volume of the {\it i}th
$k$-space shell in which $P(k)$ is estimated.

%%%%%%%%%%%%%%%%%%%%%%%%%%%%%%%%%%%%%%%%%%%%%%%%%%%%%%%
\vspace{0.2cm}
\noindent
{\bf{\em Comparison of weighting schemes}} --- The failure of the FKP
scheme to characterise the true clustering strengths of galaxies means
that it is a biased and sub-optimal estimator. We will now show
explicitly, under the assumption that the SM15 description of the
galaxy population is the correct one, that both the FKP and PVP
weighting schemes do indeed lead to sub-optimal measurements of
$P(k)$.  The weighting schemes are:

%%%%%%%%%%%%%%%%%%%%%%%%%%%%%%%%%%%%%%%%%%%%%%%%%%%%%%%
\vspace{0.1cm}
\noindent $\bullet$ {\bf The FKP weights:} These depend only on the
position of the galaxy in the survey:
\be
w_{\fkp}(\br) = 1/\left[1+\nbarg(\br)P(k)\right] \ ,
\ee
where $\nbarg(\br)$ is the mean number density of galaxies. 

%%%%%%%%%%%%%%%%%%%%%%%%%%%%%%%%%%%%%%%%%%%%%%%%%%%%%%%
\vspace{0.2cm}
\noindent $\bullet$ {\bf The PVP weights:} These depend explicitly on
the luminosity dependence of the galaxy bias and also the position in
the survey:
\be
w_{\pvp}(\br, L) = b(L)/\left[1+ \nbarg(\br) \,\bLSqbar(\br)P(k)\right] ,
\label{eq:weightsPVP}
\ee
where the luminosity-dependent galaxy bias is $b(L)\equiv\int dM
\nbar(M) b(M) N_{\g}^{(1)}(M)
\Phi(L|M)/\Phi(L)$. $\overline{b_{L}^2}(\br)\equiv\int_{\Lmin(\br)} dL
\, b^2(L)\Phi(L) /\nbarg(\br)$ is the average square of the luminosity
bias. The galaxy luminosity function is given by $\Phi(L) \equiv \int
dM \nbar(M) N_{\g}^{(1)}(M) \Phi(L|M)$, and $N^{(1)}_\g(M)$ was
introduced after \Eqn{eq:G}.

%%%%%%%%%%%%%%%%%%%%%%%%%%%%%%%%%%%%%%%%%%%%%%%%%%%%%%%
\vspace{0.2cm}
\noindent $\bullet$ {\bf Optimal weights:} In the large-scale limit,
these weights depend only on the galaxy's spatial position and its
host halo mass, and not explicitly on its luminosity. The weights are:
\be
w_{\opt}(\br, M)=b(M)/\left[1+R(M)S(\br,M)\right]\left[1+\nbar_{\rm eff}(\br) P(k)\right]  ,
\ee
where $R(M)\equiv N_\g^{(2)}(M)/N^{(1)}_\g(M)$ is the ratio of the
second and first factorial moments of the halo occupation distribution
and we introduced the effective number density of galaxies: $
\nbar_{\rm eff}(\br) \equiv \int dM \nbar(M) b^2(M)
N^{(1)}_{\g}(M){\mathcal S}(\br,M)/\left[1+R(M){\mathcal
    S}(\br,M)\right]$.  We defined $\mathcal{S}(\br, M) \equiv
\int^{\infty}_{\Lmin (\br)} dL \Phi(L|M)$ as the fraction of galaxies
hosted by haloes of mass $M$ that are observable at a spatial position
$\br$, with $\Lmin(\br)$ the minimum luminosity that a galaxy could
have and still be observable given the survey flux-limit. Explicitly,
$\Lmin(\br) = 10^{-\frac{2}{5}\left(m_{\rm lim} -25
  -M_{\odot}\right)}h^{-2}\left[d_{\rm L}(\br)/1\Mpc\right]^{-2}
[L_{\odot}]$, where $m_{\rm lim}$ is the apparent magnitude limit of
the survey, $M_{\odot}$ is the absolute magnitude of the sun, $h$ is
the dimensionless Hubble parameter and $d_{\rm L}(\br)=(1+z)\chi(z)$
is the luminosity distance in flat cosmological models. Note that
$\mathcal{S}({\bf 0}, M)=1$ and $\mathcal{S}({\bf \infty}, M)=0$.  For
more details on the $\SN$ expressions for the three weights
considered, we refer the interested reader to SM15.
%%%%%%%%%%%%%%%%%%%%%%%%%%%%%%%%%%%%%%%%%%%%%%%%%%%%%%%
%%%%%%%%%%%%%%%%%%%%%%%%%%%%%%%%%%%%%%%%%%%%%%%%%%%%%%%
\begin{figure}
  \centering{
    \includegraphics[width=8cm,clip=]{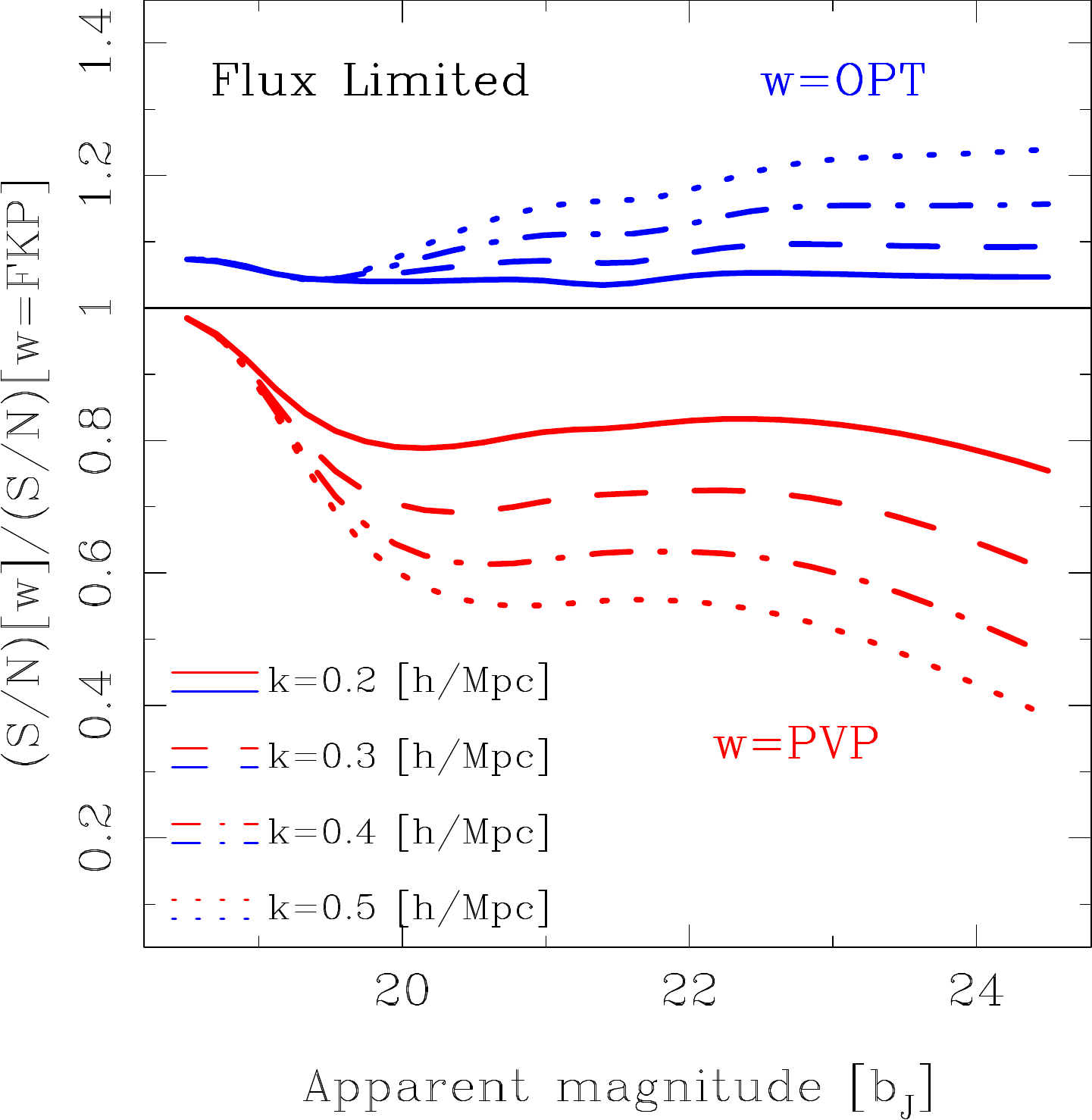}} 
\caption{\footnotesize{$\SN$ level for the optimal (blue lines, ratio$>1$)
    and the PVP (red lines, ratio $<1$) weighting schemes relative to
    the FKP one, as a function of survey $b_{\rm J}$ flux-limit. The
    solid-dashed, dot-dashed and dotted line styles denote results for
    $k\in\{0.2,\,0.3,\,0.4,\,0.5\}\kMpc$, respectively. The optimal
    scheme of SM15 clearly maximises the $\SN$ ratio for the future
    survey cases considered.}
\label{fig:SNVFluxLim}} 
\end{figure}
%%%%%%%%%%%%%%%%%%%%%%%%%%%%%%%%%%%%%%%%%%%%%%%%%%%%%%%
%%%%%%%%%%%%%%%%%%%%%%%%%%%%%%%%%%%%%%%%%%%%%%%%%%%%%%%
\begin{figure*}
\centering{
  \includegraphics[width=9cm,angle=-90,clip=]{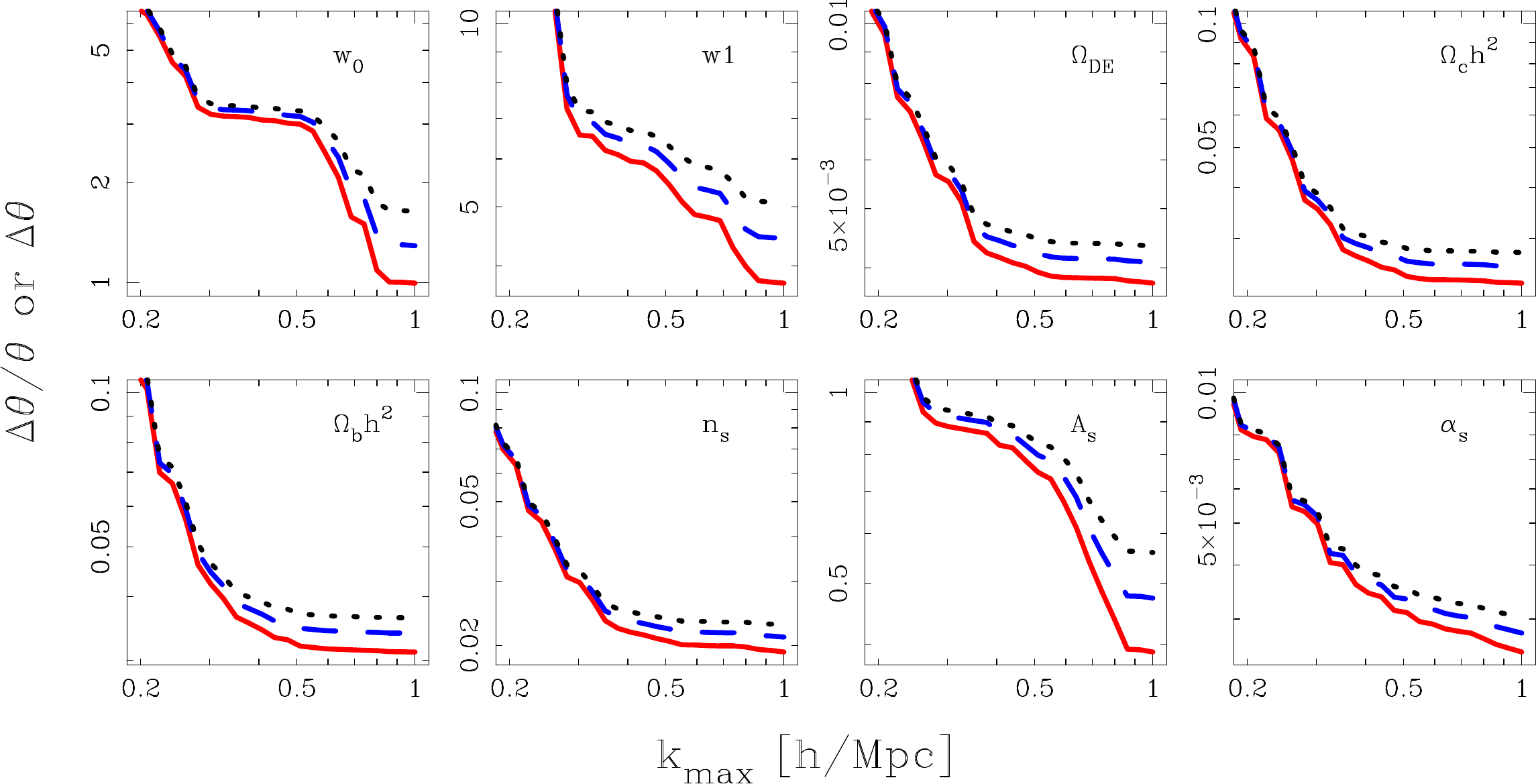}}
\caption{\footnotesize{Forecasted 1D marginalized errors and relative
    errors on cosmological parameters as a function of the maximum
    wavenumber considered in the power spectrum estimates from a
    full-sky galaxy clustering survey of depth $b_{\rm J}\sim 22$. The
    solid red, dashed blue and black dotted lines represent the SM15,
    FKP and PVP weighting schemes, respectively. The panels, going
    clockwise from the top-left show the results for the eight
    cosmological parameters considered. The largest potential
    information gains to be had from optimal weighting are in the
    measurements of $\{w_0,\,w_a,\,A_s\}$. Note that we have not
    properly taken into account the growth evolution of structure, and
    used power spectrum derivatives suitable for only a single
    redshift. }
\label{fig:1DErr}}
\end{figure*}
%%%%%%%%%%%%%%%%%%%%%%%%%%%%%%%%%%%%%%%%%%%%%%%%%%%%%%%
%%%%%%%%%%%%%%%%%%%%%%%%%%%%%%%%%%%%%%%%%%%%%%%%%%%%%%%
\begin{figure*}
\centering{
  \includegraphics[width=14cm,clip=]{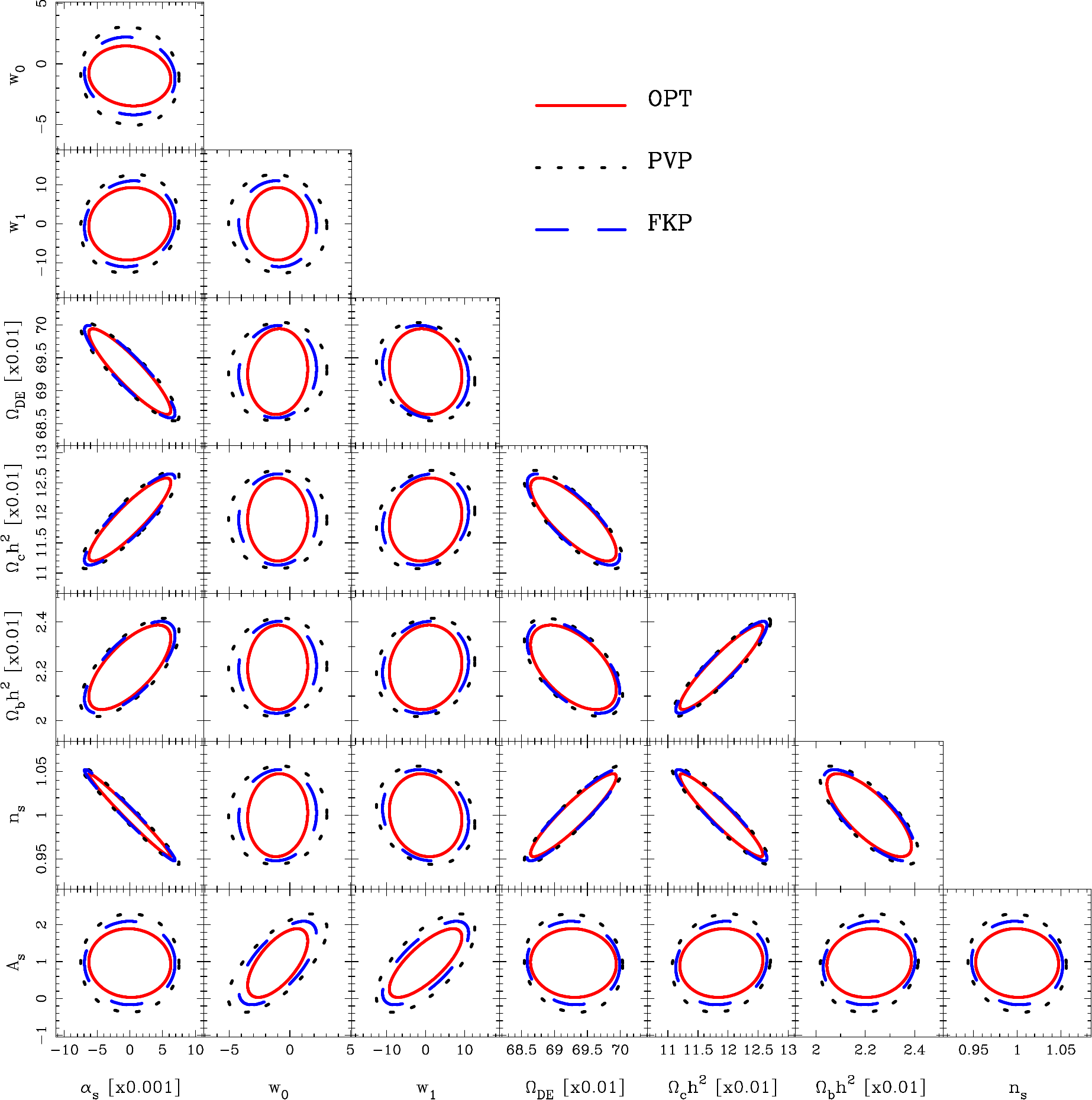}}
\caption{\footnotesize{Forecasted 2D marginalized errors on cosmological
    parameters for the eight cosmological parameters considered. The
    maximum wavenumber was set to $k=0.5\kMpc$ and the flux-limit was
    taken to be $b_{\rm J}=22$. Note that we have not properly taken
    into account the growth evolution of structure, and have used only
    power spectrum derivatives suitable for a single
    redshift. Nevertheless, it can be clearly seen that the optimal
    weighting scheme provides the tightest constraints on
    parameters.}\label{fig:2DErr}}
\end{figure*}
%%%%%%%%%%%%%%%%%%%%%%%%%%%%%%%%%%%%%%%%%%%%%%%%%%%%%%%
%%%%%%%%%%%%%%%%%%%%%%%%%%%%%%%%%%%%%%%%%%%%%%%%%%%%%%%

We now show the $\SN$ on the galaxy power spectrum corresponding to
the FKP, PVP and SM15 methods for weighting the galaxy
distribution. As a concrete example we consider a flux-limited,
full-sky galaxy redshift survey spanning the redshift range
$z=0.3$--$0.9$.  In order to evaluate the above expressions we need to
specify several model ingredients. For the evolution of $\nbar(M)$ and
$b(M)$ we use the models of \citet{ShethTormen1999}.  For the
conditional probability distribution $\Phi(L|M)$ and the first
factorial moment of the HOD $N^{(1)}_{\g}(M)$, we use the Conditional
Luminosity Function (CLF) model of \citet{Yangetal2003}. For the
second factorial moment we use the model:
$N^{(2)}_\g(M)=\beta(M)\left[N^{(1)}_\g(M)\right]^2$, where from
fitting to semi-analytic models of galaxy formation
$\beta^{1/2}(M)=1/2\log_{10}\left(M/10^{11}h^{-1}\Msol\right)$ for the
case that $M<10^{13}\Msol$ and unity otherwise
\citep{CooraySheth2002}. From these ingredients all required variables
may be computed.

Figure~\ref{fig:SNVFluxLim} shows the $\SN$ for the SM15 (blue lines)
and PVP (red lines) schemes ratioed with the $\SN$ for the FKP scheme,
respectively. The results are presented as a function of limiting
$b_{\rm J}$ magnitude and for various $k$-mode bins. Clearly, the
optimal scheme of SM15 does indeed lead to the largest $\SN$:
$\gtrsim5\%$ improvement over FKP at $k\sim0.2\kMpc$, and
$\gtrsim20\%$ improvement at $k\sim0.5\kMpc$ for surveys with depth
$b_{\rm J}\gtrsim22$. Interestingly, the scheme of PVP leads to the
least optimal set of estimates, being $\sim20\%$ lower than FKP at
$k=0.2\kMpc$ and $\sim40\%$ lower by $k=0.5\kMpc$, again for surveys
with $b_{\rm J}\gtrsim22$.

%%%%%%%%%%%%%%%%%%%%%%%%%%%%%%%%%%%%%%%%%%%%%%%%%%%%%%%
\vspace{0.2cm}
\noindent
{\bf{\em Forecasting cosmological information}} --- The
ability of a set of power spectrum band-power estimates to constrain
the cosmological parameters $\theta_{\alpha}$, can be forecasted
through construction of the Fisher information matrix
\citep{Tegmarketal1997}. For a continuum limit of Fourier modes the
Fisher matrix can be expressed as \citep{Tegmark1997}:
\be 
{\mathcal F}_{\alpha\beta} =  
\int\hspace{-0.1cm} \frac{\dk}{V(k)}
\frac{\partial \log P(k)}{\partial \theta_\alpha}
\frac{\partial \log P(k)}{\partial \theta_\beta} 
\left(\frac{{\mathcal S}}{{\mathcal N}}\right)^2\!\!(k) \ .
\label{eq:Fisher}\hspace{2cm}
\ee
Thus, in order to compute the Fisher matrix, one needs to specify the
$\SN$, and the derivatives of the power spectra with respect to the
cosmological parameters. The former were computed in the previous
section, and we estimate the latter at a single redshift. Therefore
our forecasts will be pessimistic, since we do not fully take into
account the information in the growth of structure, but here we are
only interested in the relative differences between the three
weighting schemes.

For our fiducial model we adopt a flat, dark-energy dominated
cosmological model, characterised by eight parameters:
$\theta_{\alpha}\in\{w_0,w_1,\Omega_{\rm DE},\Omega_{\rm
  c}h^2,\Omega_{\rm b}h^2,A_{\rm s},n_{\rm s},\alpha_{\rm s}\}$. The
first two characterise the equation of state for dark energy:
$w(a)=p_w/\rho_w=w_0+(1-a)w_1$; $\Omega_{\rm DE}$ is the dark energy
density parameter; $\Omega_{\rm c}h^2$ and $\Omega_{\rm b}h^2$ are the
physical densities in CDM and baryons, respectively; and $A_{\rm s}$,
$n_{\rm s}$, and $\alpha_{\rm s}$ denote the amplitude, spectral
index, and running of the primordial scalar power spectrum,
respectively. We adopt the values $\theta_{\alpha}=\{-1, 0, 0.69,
0.12, 0.02, 2.15\times10^{-9}, 0.96, 0\}$, consistent with Planck data
\citep{Planck2014XVI}. The power spectrum derivatives we compute
through finite differencing matter power spectra from CAMB
\citep{Lewisetal2000}.

Figure~\ref{fig:1DErr} shows the forecasted 1D marginalized errors on
the parameters, as a function of the maximum wavenumber $k_{\rm max}$
entering the integral of \Eqn{eq:Fisher}. The panels show the
fractional error, or if the fiducial value is zero, the
error. Clearly, the smallest errors are obtained when one implements
the optimal weighting scheme of SM15 (red solid lines), followed by
FKP (blue dashed line) and then PVP (black dotted lines).  We notice
that the constraints on $(A_{\rm s},w_0,w_1)$ show the most
significant improvements from the optimal weighting.

Figure~\ref{fig:2DErr} shows the forecasted 2D marginalized errors on
various parameter combinations. The line styles are the same as in
Figure~\ref{fig:1DErr}. Again, the optimal weighting of SM15 performs
best and the parameters $(A_{\rm s},w_0,w_1)$ appear to be the most
affected by the new scheme.

%%%%%%%%%%%%%%%%%%%%%%%%%%%%%%%%%%%%%%%%%%%%%%%%%%%%%%%
\vspace{0.2cm}
\noindent
{\bf{\em Conclusions}} --- In this letter we presented an overview of
the optimal power spectrum estimation scheme of SM15. We argued that
the FKP scheme was biased and sub-optimal since it does not take into
account variations of clustering with the galaxy sample. We argued
that the SM15 framework, which encodes several key concepts from the
theory of galaxy formation, is able to describe these variations. We
evaluated the $\SN$ resulting from the FKP, PVP, and SM15 weighting
schemes for the case of an all-sky galaxy survey. The SM15 weighting
scheme was found to indeed be the most efficient estimator. We then
turned to the issue of cosmological information and using the Fisher
matrix approach showed that the SM15 scheme also produced the smallest
errors on cosmological parameters. In particular, the parameters
governing the amplitude of the primordial power spectrum and the
evolution of the dark energy equation of state were noticeably
improved.

In the previous studies of \citep{Seljaketal2009,Hamausetal2010} it
was claimed that weighting galaxy groups by some linear function of
the halo mass would lead to reduced shot-noise and hence boosted
signal-to-noise estimates.  In the limit of large number of galaxies
per halo, the mass dependence of the SM15 weights is $w\propto
b(M)/N_{\g}^{(1)}(M)$, whereas in the limit of small numbers $w\propto
b(M)$. Clearly, the SM15 weighting scheme does not follow this mass
scaling. The calculation of SM15 has maximised the $\SN$ on power
spectrum estimates, albeit under certain assumptions, whereas the
calculations of \citet{Seljaketal2009} and later
\citet{Hamausetal2010} have minimised the stochasticity on the halo
density field. These two things are not obviously the same.  We argue
that our approach is the correct path to follow since, by design, it
minimises directly errors in the power spectrum and it has clearly
built into its framework the corner stones of galaxy formation theory.
%%%%%%%%%%%%%%%%%%%%%%%%%%%%%%%%%%%%%%%%%%%%%%%%%%%%%%%
\bibliographystyle{apsrev.bst} 
%\bibliography{refs.bib}

\end{document}